\documentclass[aps,pra,preprint,superscriptaddress,nofootinbib]{revtex4-2}

\usepackage{graphicx}
\usepackage{mathtools}
\usepackage{amssymb}

\usepackage{verbatim}
\usepackage{color}
\usepackage{amsthm}

\usepackage{enumitem}
\usepackage{booktabs}
\usepackage{algorithm,algorithmicx,algpseudocode}

\usepackage{aliascnt}
\usepackage{hyperref}
\usepackage{cleveref}
\usepackage{tikz-cd}

\newtheorem{theorem}{Theorem}

\newaliascnt{definition}{theorem}
\newtheorem{definition}[definition]{Definition}
\aliascntresetthe{definition}

\newaliascnt{lemma}{theorem}
\newtheorem{lemma}[lemma]{Lemma}
\aliascntresetthe{lemma}

\newaliascnt{proposition}{theorem}
\newtheorem{proposition}[proposition]{Proposition}
\aliascntresetthe{proposition}

\newaliascnt{corollary}{theorem}
\newtheorem{corollary}[corollary]{Corollary}
\aliascntresetthe{corollary}

\newaliascnt{remark}{theorem}
\theoremstyle{remark}
\newtheorem{remark}[remark]{Remark}
\aliascntresetthe{remark}

\newcommand{\mycell}[2]{\parbox[t]{#1}{\raggedright #2}}

\newcommand{\abs}[1]{\left\vert#1\right\vert}
\newcommand{\defeq}{\vcentcolon=}

\newcommand{\tr}[1]{ \mathrm{tr}\left(#1\right) }

\newcommand{\ket}[1]{\vert#1\rangle}
\newcommand{\Zq}{\mathbb{Z} / q \mathbb{Z}}
\newcommand{\Ztwo}{\mathbb{Z} / 2 \mathbb{Z}}

\newcommand{\Zfour}{\mathbb{Z} / 4 \mathbb{Z}}
\newcommand{\CG}{\mathbb{C} \left[ G \right]}
\newcommand{\spanC}{\mathrm{span}_\mathbb{C}}

\DeclareMathOperator{\End}{End}

\begin{document}

\title{One-Query Quantum Algorithms for the Index-$q$ Hidden Subgroup Problem}
\thanks{\textcopyright\ 2026 American Physical Society.
	Accepted for publication in Physical Review A.}

\author{Amit Te'eni}
\email[Corresponding author, ]{amit.teeni@biu.ac.il}
\affiliation{Faculty of Engineering and the Institute of Nanotechnology and Advanced Materials, Bar-Ilan University, Ramat Gan 5290002, Israel}

\author{Yaron Oz}
\affiliation{School of Physics and Astronomy, Tel-Aviv University, Tel-Aviv, 69978 Israel}

\author{Eliahu Cohen}
\affiliation{Faculty of Engineering and the Institute of Nanotechnology and Advanced Materials, Bar-Ilan University, Ramat Gan 5290002, Israel}
\affiliation{Institute for Quantum Studies, Chapman University, Orange, California 92866, USA}

	
	\begin{abstract}
		The quantum Fourier transform (QFT) is central to many quantum algorithms, yet its necessity is not always well understood. We re-examine its role in canonical query problems. The Deutsch--Jozsa algorithm requires neither a QFT nor a domain group structure. In contrast, the Bernstein--Vazirani problem is an instance of the hidden subgroup problem (HSP), where the hidden subgroup has either index $1$ or $2$, and the Bernstein--Vazirani algorithm exploits this promise to solve the problem with a single query. Motivated by these insights, we introduce the minimal-label index-$q$ HSP: given an oracle with a $q$-element output alphabet, determine whether the hidden subgroup $H \le G$ has index $1$ or $q$, and, when possible, identify $H$. We present a single-query algorithm that always distinguishes index $1$ from $q$.
		For finite abelian groups, the same algorithm exactly identifies $H$ under explicit necessary and sufficient conditions: $G/H$ is cyclic, and the unknown coset-to-label map satisfies a certain Fourier-support condition. We further extend this identification algorithm to normal subgroups of arbitrary finite groups, with cyclic quotient. When $H$ is normal and $q$ is prime, the cyclicity and Fourier-support conditions hold automatically, yielding unconditional single-query recovery.
		In contrast, the Shor--Kitaev sampling approach cannot guarantee exact recovery from a single sample. Our results sharpen the landscape of one-query quantum solvability for HSPs.
	\end{abstract}
	
	\maketitle
	
	\section{Introduction}\label{sec:intro}
	Quantum algorithms often achieve their power by exploiting algebraic structures. The quantum Fourier transform (QFT), in particular, underlies several landmark results, including Shor's factoring algorithm~\cite{shor1994algorithms,Shor1997}, Simon's algorithm~\cite{Simon1994Power}, and the more general hidden subgroup framework~\cite{kitaev1995quantummeasurementsabelianstabilizer,Jozsa1998Quantum,Mosca1999Quantum,Ettinger1999Quantum,Jozsa2001Quantum,lomonaco2002quantumhiddensubgroupproblems,lomont2004hiddensubgroupproblem,Kuperberg2005Subexponential,moore2020hidden,kuperberg2025hiddensubgroupprobleminfinite} (see also~\cite{nielsen2010quantum} for a pedagogical exposition). Even in simpler settings, such as the Deutsch--Jozsa~\cite{Deutsch1992} and Bernstein--Vazirani~\cite{Bernstein1997} algorithms, the QFT appears in the guise of the Hadamard transform (QFT over $ \left( \mathbb{Z} / 2 \mathbb{Z} \right)^n$). While this formalism has proven both elegant and powerful, the prominent role of the Fourier transform raises a natural question: to what extent is the QFT truly essential to the operation of these algorithms, and to what extent is it simply a convenient mathematical description?
	
	Standard circuit-based descriptions of oracle algorithms usually assume that the computer is built from qubits, with unitaries decomposed into single- and two-qubit gates. This perspective is natural from an implementation standpoint, but it can obscure the actual mathematical structure that accounts for the algorithm's success. Our aim is to separate these two layers---the algebraic structure that drives the computational advantage, and the gate-level description in terms of qubits---in order to better understand when Fourier analysis is indispensable and when more general unitary operations suffice. Here we focus on oracle problems, where the task is to classify an unknown function $f$ using as few queries as possible. 
	
	For example, the standard description of the Deutsch--Jozsa (DJ) algorithm applies a Hadamard gate on all $n$ qubits. The resulting unitary, $ F_{\left( \Ztwo \right)^n} \defeq H^{\otimes n} $, also referred to as the \textit{Hadamard transform}, is the quantum Fourier transform (QFT) associated with the group $ \left( \Ztwo \right)^n $. However, this operator can be replaced by any unitary that maps $\ket{0}$ to the ``democratic superposition'' state~\cite{hoyer1999conjugated}. This observation is founded on the insight that no particular group structure on the domain set $ \left\{ 0,1 \right\}^n $ plays any role in the formulation of the Deutsch--Jozsa problem~\cite{vicary2013topological}; moreover, the same observation is key for several generalizations of the Deutsch--Jozsa problem and algorithm~\cite{batty2006extending,vicary2013topological}.
	
	While the Deutsch--Jozsa algorithm achieves exact (zero-error) success using only this coarse promise, a parallel line of research has sought to characterize exact one-query algorithms for more highly structured problems~\cite{chen2020Characterization, xu2021partial, qiu2021exact, ye2023characterization}. Moreover, significant effort has been devoted to designing deterministic quantum algorithms for the hidden subgroup problem (HSP) and its generalizations~\cite{brassard1997exact, mosca2004exact, childs2010quantum, friedl2014hidden, ye2021query, ye2022deterministic}. A natural question uniting these two directions is whether we can pinpoint the exact structural promises that enable single-query identification of hidden subgroups, without relying on the probabilistic sampling inherent to standard approaches like the Shor--Kitaev algorithm.
	
	To resolve this, we first distill the Deutsch--Jozsa problem to its core, clarifying which sets do and do not carry an algebraic structure. This perspective allows us to bridge the setting of Deutsch--Jozsa with the algebraic framework of the HSP, introducing the \emph{minimal-label index-$q$ hidden subgroup problem}. This is a version of the HSP where the output alphabet is $ \Zq $, and the hidden subgroup $ H \le G $ is promised to have index either $1$ or $q$. We show a single-query algorithm that \textit{always} decides whether the index is $1$ or $q$.
	For finite abelian groups, we then identify necessary and sufficient conditions under which the same single query \textit{exactly} recovers $H$: namely, $G/H$ is cyclic, and the unknown coset-labeling map $\iota$ satisfies a Fourier-support condition (which translates geometrically to a combinatorial \emph{coset-balance} requirement). We further extend this identification algorithm to non-abelian groups $G$ where the hidden subgroup $H$ is promised to be normal with cyclic quotient.
	If $q$ is prime, the quotient $G/H$ is automatically cyclic, and we prove that the Fourier-support condition holds for any map $\iota$, yielding unconditional recovery with a single query. In contrast, we prove that a single Shor--Kitaev sample cannot guarantee identification (success probability $\varphi \left( q \right) / q$ at best).
	Finally, we show that the Bernstein--Vazirani (BV) problem can be cast as a minimal-label index-$2$ HSP on $G= \left( \mathbb{Z} / 2\mathbb{Z} \right)^n $ with $f \left( x \right) = \vec{x} \cdot \vec{s} \mod 2 $ and $H = \ker f$; and the BV algorithm is a special case of our generic single-query algorithm for the minimal-label index-$q$ HSP.
	Thus, our work clarifies the role of the QFT across several canonical algorithms, sharpening the boundary between DJ, generic abelian-HSP methods, and BV. The results are summarized in \Cref{tab:onequery}.
	
	\begin{table*}
		\caption{Comparison of one-query quantum procedures. The phase and shift oracles are defined in \Cref{sec:trick}. The following existing algorithms are discussed in the following sections: SK in \Cref{sec:HSP}; DJH (which generalizes Deutsch--Jozsa) in \Cref{sec:DJH}; and BV in \Cref{sec:BV}. The minimal-label index-$q$ HSP and the single-query algorithm that solves it are introduced in \Cref{sec:idx2HSP}. The Fourier-support condition is defined in \Cref{def:Fourier-support}.}
		\label{tab:onequery}
		\scriptsize
		\begin{tabular}{ l l l l l l}
			\toprule
			\mycell{0.14\textwidth}{\textbf{Problem / algorithm}} &
			\mycell{0.16\textwidth}{\textbf{Group / hidden subgroup / quotient}} &
			\mycell{0.14\textwidth}{\textbf{Oracle \& Output Structure}} &
			\mycell{0.14\textwidth}{\textbf{Oracle model}} &
			\mycell{0.2\textwidth}{\textbf{$1$-query deterministic?}} &
			\mycell{0.18\textwidth}{\textbf{Role of QFT}}\\
			\midrule
			
			\mycell{0.14\textwidth}{\textbf{Deutsch--Jozsa--H{\o}yer (DJH)} (balanced vs.\ constant)} &
			\mycell{0.16\textwidth}{N/A (no structure assumed on oracle domain)} &
			\mycell{0.14\textwidth}{Abelian group $G$} &
			\mycell{0.14\textwidth}{Phase, w.r.t. any nontrivial $\chi \in \hat{G}$} &
			\mycell{0.2\textwidth}{Yes (decision)} &
			\mycell{0.18\textwidth}{Incidental (any unitary with uniform first column)} \\
			\addlinespace
			
			\mycell{0.14\textwidth}{\textbf{Bernstein--Vazirani (BV)}} &
			\mycell{0.16\textwidth}{$G= \left( \mathbb{Z} / 2 \mathbb{Z} \right)^n$, $H=\ker \left( x\mapsto \vec{s}\cdot \vec{x} \right)$, so $G/H \cong \mathbb{Z} / 2 \mathbb{Z}$ or trivial} &
			\mycell{0.14\textwidth}{$\mathbb{Z} / 2 \mathbb{Z}$} &
			\mycell{0.14\textwidth}{Phase, w.r.t. nontrivial $\chi \in \widehat{\mathbb{Z} / 2 \mathbb{Z}}$} &
			\mycell{0.2\textwidth}{Yes} &
			\mycell{0.18\textwidth}{Essential: QFT over $\left( \mathbb{Z} / 2 \mathbb{Z} \right)^n$} \\
			\addlinespace
			
			\mycell{0.14\textwidth}{\textbf{Minimal-label index-$q$ HSP: Decision} (does $[G:H]=1$ or $q$?)} &
			\mycell{0.16\textwidth}{$G$ finite, $ \left[ G : H \right] \in \{1, q\}$} &
			\mycell{0.14\textwidth}{$|X|=q$ with fixed $\mathbb{Z} / q \mathbb{Z}$ group law} &
			\mycell{0.14\textwidth}{Shift} &
			\mycell{0.2\textwidth}{Yes} &
			\mycell{0.18\textwidth}{Incidental for $G$ (any uniform-column matrix); QFT over $X$ used for shift-to-phase conversion} \\
			\addlinespace
			
			\mycell{0.14\textwidth}{\textbf{Minimal-label index-$q$ HSP: Identification}} &
			\mycell{0.16\textwidth}{$G$ finite, $H \trianglelefteq G$, $ G / H \cong \mathbb{Z} / q \mathbb{Z} $ or trivial} &
			\mycell{0.14\textwidth}{$|X|=q$ with fixed $\mathbb{Z} / q \mathbb{Z}$ group law} &
			\mycell{0.14\textwidth}{Shift} &
			\mycell{0.2\textwidth}{Yes unconditionally for prime $q$; for composite $q$, iff unknown bijection $\iota$ satisfies Fourier-support condition} &
			\mycell{0.18\textwidth}{Essential: QFT over $G$ in presented implementation (QFT over $ G / [G,G] $ also suffices, see \Cref{rmk:abelianization})} \\
			\addlinespace
			
			\mycell{0.14\textwidth}{\textbf{Abelian HSP / Shor--Kitaev (SK)}} &
			\mycell{0.16\textwidth}{$G$ finite abelian, $H \le G$ arbitrary} &
			\mycell{0.14\textwidth}{None} &
			\mycell{0.14\textwidth}{Shift} &
			\mycell{0.2\textwidth}{No} &
			\mycell{0.18\textwidth}{Essential: QFT over $G$} \\
			\bottomrule
		\end{tabular}
	\end{table*}
	
	This paper is structured as follows. In \Cref{sec:QFT} we review the quantum Fourier transform and present two of its most common applications: the implementation of phase oracles using shift oracles, and the Shor--Kitaev algorithm. In \Cref{sec:DJ} we discuss the Deutsch--Jozsa problem and algorithm, showing that any unitary $V$ whose first column has equal entries can be used as the pre-oracle unitary, with $ V^\dagger $ as the post-oracle unitary, without further modification to the standard algorithm. We also describe several known generalizations of the Deutsch--Jozsa problem. The core of the paper is \Cref{sec:idx2HSP}, where we introduce the index-$q$ HSP and a single-query algorithm that solves it with certainty. We illustrate the problem and algorithm with a concrete example, which can be viewed as a refinement of Deutsch--Jozsa with $n=2$. In \Cref{sec:BV} we examine the Bernstein--Vazirani problem and algorithm, showing that they constitute a special case of the index-$2$ HSP, and mention several existing generalizations. We conclude in \Cref{sec:conclusion} with a summary of our results and a discussion of directions for future work.
	
	\section{Common applications of the quantum Fourier transform}\label{sec:QFT}
	In this section we briefly review the QFT and two of its common applications in quantum algorithms. Throughout this paper, let $ \mathbb{C}A $ be the Hilbert space with orthonormal basis corresponding to elements of the set $A$. For a finite abelian group $G$, the QFT comprises a distinguished unitary operator on $ \mathbb{C}G $. 
	This space carries the regular representation of $G$, which decomposes into a direct sum of all one-dimensional representations (characters). Thus, $ \mathbb{C}G $ has two natural orthonormal bases: $ \left\{ \ket{ g } \right\}_{g \in G} $ (the computational basis) and $ \left\{ \ket{ \hat{\delta}_\chi} \right\}_{\chi \in \hat{G}} $ (the character basis), where $ \hat{G} $ is the set of all characters of $G$, and $ \ket{ \hat{\delta}_\chi} $ spans the representation defined by $\chi$.
	The QFT transforms a vector written in the computational basis to one written in the character basis.
	
	Let us elaborate further. The complex irreducible representations (irreps) of a finite abelian group $G$ are all one-dimensional, i.e. characters $ \chi : G \rightarrow \mathrm{U} ( 1 ) $. The set $\hat{G} $ of all characters of $G$ is endowed with pointwise multiplication: $ \left( \chi_1 \chi_2 \right) ( g ) \defeq \chi_1 ( g ) \chi_2 ( g )$. $\hat{G}$ with this product forms a group---the dual group of $G$. $G$ and $\hat{G}$ are isomorphic, but generally there is no canonical isomorphism. As we mentioned, $ \mathbb{C}G $ carries the regular representation of $G$---an element $ a \in G $ acts on basis vectors as follows:
	\begin{equation}\label{shift_a}
		\forall a \in G, \quad a \cdot \ket{g} \defeq \ket{a \oplus g} ,
	\end{equation}
	where $ \oplus $ denotes the group law of $G$. The regular representation decomposes into a direct sum of all characters, i.e.,
	\begin{equation}\label{iso_of_reps}
		\mathbb{C}G \cong \bigoplus_{\chi \in \hat{G}} V_\chi ,
	\end{equation}
	where $ V_\chi $ is a one-dimensional subspace with the representation $ a \cdot \ket{v} = \chi ( a ) \ket{v} $ for all $ a \in G, \ket{v} \in V_\chi $. We can construct $V_\chi$ explicitly:
	\begin{equation}
		V_\chi \defeq \mathrm{span}_{\mathbb{C}} \left\{ \ket{ \hat{\delta}_\chi} \right\} , \quad \ket{ \hat{\delta}_\chi} \defeq \frac{1}{\sqrt{\abs{G}}} \sum_{g \in G} \chi^* ( g ) \ket{g} .
	\end{equation}
	Let us see that $ \ket{ \hat{\delta}_\chi} $ indeed carries the $\chi$ representation:
	\begin{align}\label{delta_chi_spans_char_chi}
		\forall a \in G, \quad a \cdot \ket{ \hat{\delta}_\chi} & = \frac{1}{\sqrt{\abs{G}}} \sum_{g \in G} \chi^* ( g ) \ket{a \oplus g} = \frac{1}{\sqrt{\abs{G}}} \sum_{h \in G} \chi^* \left( h \oplus \left( -a \right) \right) \ket{h} = \nonumber\\
		& = \chi ( a ) \frac{1}{\sqrt{\abs{G}}} \sum_{h \in G} \chi^* ( h ) \ket{h} = \chi ( a ) \ket{ \hat{\delta}_\chi} .
	\end{align}
	Using orthogonality of characters, one can show that $ \left\{ \ket{ \hat{\delta}_\chi} \right\}_{\chi \in \hat{G} } $ is an orthonormal basis. Now, let $ \ket{\psi} = \sum_{g \in G} a_g \ket{g} $ be an arbitrary state; equivalently, we may think of $ \ket{\psi} $ as a function $ G \rightarrow \mathbb{C} $. We know $ \ket{\psi} $ can also be written as $ \ket{\psi} = \sum_{\chi \in \hat{G}} b_\chi \ket{ \hat{\delta}_\chi} $ (a function $ \hat{G} \rightarrow \mathbb{C} $). We wish to define the QFT as the unitary $F$ that translates the former perspective to the latter, i.e.
	\begin{equation*}
		F \begin{pmatrix}
			\vdots \\
			a_g \\
			\vdots
		\end{pmatrix} = \begin{pmatrix}
			\vdots \\
			b_\chi \\
			\vdots
		\end{pmatrix} .
	\end{equation*}
	Thus, the columns of $F$ are labeled by the elements of $G$; its rows---by the elements of $ \hat{G} $; and 
	\begin{equation}
		\ket{g} = \sum_{ \chi \in \hat{G} } F_{ \chi g } \ket{ \hat{\delta}_\chi} \quad \Rightarrow \quad F_{ \chi g } = \left\langle \hat{\delta}_\chi \vert g \right\rangle = \frac{ \chi ( g ) }{ \sqrt{\abs{G}} } .
	\end{equation}
	Hence, the matrix $ F $ is simply the character table of $G$ divided by $ \sqrt{\abs{G}} $.
	
	The Fourier-transformed vector $ F \ket{\psi} $ is written in terms of ``another'' computational basis, which we denote $ \left\{ \ket{ \chi } \right\}_{\chi \in \hat{G}} $. We define $F$ explicitly:
	\begin{equation}
		F \ket{ g } = \frac{1}{ \sqrt{\abs{G}} } \sum_{ \chi \in \hat{G} } \chi ( g ) \ket{\chi} .
	\end{equation}
	Thus, the QFT can be seen as implementing a canonical unitary isomorphism $ \mathbb{C}G \cong \mathbb{C}\hat{G} $. On the logical level, at each phase of the algorithm we should know whether the computational basis is currently treated as $ \left\{ \ket{g} \right\}_{g \in G} $ or $ \left\{ \ket{ \chi } \right\}_{\chi \in \hat{G}} $ (i.e., if the state's components refer to $G$ or $ \hat{G} $ elements).
	
	Just like the usual Fourier transform, the quantum Fourier transform has a ``shift-to-phase'' property:
	\begin{equation}\label{phase_to_shift}
		\forall a \in G, \quad F \left( a \cdot \ket{\psi} \right) = D_a F \ket{ \psi } ,
	\end{equation}
	where $ D_a \ket{ \chi} \defeq \chi \left( a \right) \ket{ \chi } $ is the diagonal ``phase action'' of $a$ on $ \mathbb{C} \hat{G} $. \eqref{phase_to_shift} means that the Fourier transform implements the isomorphism of representations \eqref{iso_of_reps}. One can prove it for basis vectors (a straightforward computation) and then extend linearly.
	The ``shift-to-phase'' property lies at the heart of both applications of the QFT that we review in the following subsections.
	
	\subsection{The general abelian ``shift-to-phase'' trick}\label{sec:trick}
	In this subsection, we present an important (yet overlooked) application of the quantum Fourier transform---namely, implementing the phase oracle using a shift oracle.
	The treatment here follows a comment made in passing by Zhandry~\cite{zhandry2015quantum}.
	
	Let $f : S \rightarrow G $ be a function from an \textit{unstructured} finite set $S$ to a finite abelian group $G$, and let $ \chi : G \rightarrow \mathrm{U} \left( 1 \right) $ be an irrep of $G$ (hence a character). The prototypical example to keep in mind is $ G = \mathbb{Z} / 2 \mathbb{Z} $ and $\chi \left( a \right) = \left( -1 \right)^a $, where $a \in \mathbb{Z} / 2 \mathbb{Z} $ is interpreted as either $0$ or $1$ (so $\chi$ is the alternating representation).
	
	Suppose we have access to $f$ via the \textit{shift oracle}:
	\begin{equation}
		\begin{split}
			& U_f^{\mathrm{shift}} : \mathbb{C}S \otimes \mathbb{C}G \xrightarrow{\cong} \mathbb{C}S \otimes \mathbb{C}G \\
			& U_f^{\mathrm{shift}} \ket{ x } \otimes \ket{ g } = \ket{ x } \otimes \ket{ f\left( x \right) \oplus g } .
		\end{split}
	\end{equation}
	We would like to use $ U_f^{\mathrm{shift}} $ to implement the \textit{phase oracle}:
	\begin{equation}\label{phase_oracle}
		\begin{split}
			& U_f^{\mathrm{phase}} : \mathbb{C}S \xrightarrow{\cong} \mathbb{C}S \\
			& U_f^{\mathrm{phase}} \ket{ x } = \chi \left( f \left( x \right) \right) \ket{ x } .
		\end{split}
	\end{equation}
	This can be done using the quantum Fourier transform---but not with respect to $S$ (which is \textit{not} assumed to have any group structure), but rather with respect to $G$.
	
	Consider a single element $ a \in G $ (which we may think of as the image under $f$ of some $x \in S$), and consider the Hilbert space $ \mathbb{C}G $. Recall this vector space carries the regular representation of $ G $, which decomposes into a direct sum of all irreps of $G$. Intuitively, we wish to ``extract'' the copy of our distinguished irrep $\chi$.
	For example, the copy of the alternating irrep inside $ \mathbb{C} \left[ \mathbb{Z} / 2 \mathbb{Z} \right] $ is spanned by $ \ket{-} $, so a shift by $a \in \mathbb{Z} / 2 \mathbb{Z} $ on this subset is just multiplication by the phase $ \left( -1 \right)^a $. This is the most well-known version of the ``shift-to-phase'' trick.
	
	To apply the trick in the general case, we first treat the computational basis as $ \left\{ \ket{\rho} \right\}_{\rho \in \hat{G}} $. We prepare the state $ \ket{\chi} $ (the computational-basis-state corresponding to our distinguished irrep $\chi$), apply the \textit{inverse} QFT and obtain:
	\begin{equation}
		F^\dagger \ket{ \chi} =  \frac{1}{\sqrt{\abs{G}}} \sum_{g \in G} \chi^* \left( g \right) \ket{g} =  \ket{ \hat{\delta}_\chi } .
	\end{equation}
	Recall $ \ket{ \hat{\delta}_\chi } $ spans the one-dimensional irrep $\chi$, meaning that on this state, the linear ``shift-by-$a$'' operation \eqref{shift_a} has the effect of multiplication by the scalar $ \chi \left( a \right) $ \eqref{delta_chi_spans_char_chi}. One also sees this by substituting $ \ket{\chi} = F \ket{\psi} $ in \eqref{phase_to_shift}:
	\begin{equation}
		\forall a \in G, \quad a \cdot \left( F^\dagger \ket{\chi} \right) = F^\dagger D_a \ket{\chi} = \chi \left( a \right) F^\dagger \ket{\chi} .
	\end{equation}
	Thus, for any $ x \in S $, we may use $ U_f^{\mathrm{shift}} $ to shift the state $ \ket{ \hat{\delta}_\chi } $ by $ f \left( x \right) $ and obtain:
	\begin{equation}
		U_f^{\mathrm{shift}} \ket{x} \otimes \ket{ \hat{\delta}_\chi } = \chi \left( f \left( x \right) \right) \ket{x} \otimes \ket{ \hat{\delta}_\chi }
	\end{equation}
	as required.
	
	\subsection{The Shor--Kitaev algorithm}\label{sec:HSP}
	In the hidden subgroup problem, we are given a group $G$, a set $X$ and an unknown function $f : G \rightarrow X$ that \textit{hides} a subgroup $H \le G$. I.e., we have the following commutative triangle:
	\begin{equation}\label{triangle}
		\begin{tikzcd}
			G \arrow[rr,"f"] \arrow[ddr,two heads,"\pi"] && X \\ && \\
			& G/H \arrow[uur,hook,"\iota"]
		\end{tikzcd}	
	\end{equation}
	hence $f$ is the composition of the canonical projection $ \pi: G \twoheadrightarrow G/H $ with any one-to-one map (of sets) $ \iota : G/H \hookrightarrow X $. Given oracle access to $f$, we are tasked with finding the hidden subgroup $H$.
	
	For finite abelian $G$, the hidden subgroup problem can be solved efficiently using the Shor--Kitaev algorithm, which uses the quantum Fourier transform. In principle, $X$ is an unstructured set; but to define the shift oracle, we need to assume some arbitrary abelian group structure on $X$. This group law is part of the oracle specification, but no compatibility with the unknown coset labeling is assumed; in particular, $f$ need not be a group homomorphism. Thus, the QFT is performed with respect to the group $G$, rather than $X$.
	The algorithm utilizes the Hilbert space $ \mathbb{C} G \otimes \mathbb{C} X $. We now describe it:
	\begin{enumerate}\setcounter{enumi}{-1}
		\item Consider the computational basis as indexed by $\hat{G}$, and prepare $ \ket{ \tilde{\psi}_0} = \ket{\chi_0} \otimes \ket{0} $, where $ \chi_0 \in \hat{G} $ is the trivial character.
		
		\item Apply the inverse QFT $F^\dagger$ to the first register, obtaining a \textit{democratic superposition}:
		\begin{equation*}
			\ket{ \tilde{\psi}_1} = \ket{ \hat{\delta}_{\chi_0} } \otimes \ket{0} = \frac{1}{\sqrt{\abs{G}}} \sum_{g \in G} \ket{g} \otimes \ket{0} .
		\end{equation*}
		
		\item Perform a single oracle query, to obtain $ \ket{ \tilde{\psi}_2} = \frac{1}{\sqrt{\abs{G}}} \sum_{g \in G} \ket{g} \otimes \ket{ f \left( g \right) } $.
		
		\item Measure the second register in its computational basis (indexed by $X$) and discard the outcome. From the assumption on $f$, the resulting state is a sum $ \ket{ \tilde{\psi}_3 } =  \frac{1}{\sqrt{ \abs{H} }} \sum_{g \in a \oplus H} \ket{g} $ over the elements of some coset $a \oplus H$ of $H$. Equivalently, we can write it as $ \frac{1}{\sqrt{ \abs{H} }} \sum_{h \in H} \ket{ a \oplus h} $, or better yet: $\ket{ \tilde{\psi}_3 } = a \cdot \left( \frac{1}{\sqrt{ \abs{H} }} \sum_{h \in H} \ket{ h} \right) $.
		
		\item Apply the QFT $F$. From \eqref{phase_to_shift}, we obtain the state
		$ \ket{ \tilde{\psi}_4 } = \frac{1}{\sqrt{ \abs{H} }} \sum_{h \in H} D_a F \ket{h} $. We find:
		\begin{equation*}
			\ket{ \tilde{\psi}_4 } = \frac{1}{ \sqrt{ \abs{H} } } \sum_{h \in H} \frac{1}{ \sqrt{ \abs{G} } } \sum_{ \rho \in \hat{G} } \rho \left( a \right) \rho \left( h \right) \ket{\rho} = 
			\sqrt{ \frac{\abs{H}}{\abs{G}} } \sum_{ \rho \in \hat{G} } \rho \left( a \right) \left( \frac{1}{ \abs{H} } \sum_{h \in H} \rho \left( h \right) \right) \ket{\rho} .
		\end{equation*}
		The parenthesized expression is the inner product of $ \rho \vert_H $ with the trivial representation of $H$, where $ \rho \vert_H $ is the restriction of $\rho$ to $H$. Thus, we obtain $ \ket{ \tilde{\psi}_4 } = \frac{1}{ \sqrt{ \abs{ H^\perp } } } \sum_{ \rho \in H^\perp } \rho \left( a \right) \ket{ \rho } $, where $ H^\perp \le \hat{G} $ is the subgroup comprising all characters $\rho$ of $G$ that are trivial on $H$ (i.e. $ \rho \left( H \right) = 1 $).
		
		\item Measure $ \ket{ \tilde{\psi}_4 } $ in the computational basis (indexed by $\hat{G}$), and get an element of $H^\perp$ (all such elements are obtained with equal probability).
	\end{enumerate}
	If we repeat the process sufficiently many times, we can recover $ H^\perp $, and hence $H$, with high probability. For a detailed analysis of the information about $H$ obtained from $t$ queries of the Shor--Kitaev algorithm, see Section IV.C of \cite{teeni2026optimization}.
	
	\section{The Deutsch--Jozsa problem and algorithm}\label{sec:DJ}
	Let us briefly describe the Deutsch--Jozsa (DJ) problem.
	Let $N$ be an even number (for now it is unnecessary to assume $N=2^n$), and denote $ \left[ N \right] = \left\{ 0, \ldots, N-1 \right\} $. Let $f : \left[ N \right] \rightarrow \mathbb{Z} / 2 \mathbb{Z} $ be a function promised to be either constant or balanced. For the sake of simplicity, suppose that we are given access to $f$ via the phase oracle $U_f$. This oracle acts on the Hilbert space $ \mathcal{H} = \mathbb{C}^N $, and is diagonal in the standard basis:
	\begin{equation}
		\forall x \in \left[ N \right], \quad U_f \ket{x} = \left( -1 \right)^{f \left( x \right)} \ket{x} .
	\end{equation}
	We are tasked with finding whether $f$ is constant or balanced.
	
	To solve the DJ problem using a single query, it suffices to choose an $ N \times N $ unitary matrix $ V $ such that:
	\begin{equation}
		\forall x \in \left[ N \right], \quad \left\langle x \vert V \vert 0 \right\rangle = 1 / \sqrt{N} ,
	\end{equation}
	i.e. the first column of $V$ is proportional to the vector of all ones. \textbf{No further assumption on $V$ is necessary.} For example, if $N=2^n$ then we can choose $V$ to be the QFT associated with the group $ \left( \mathbb{Z} / 2 \mathbb{Z} \right)^n $, but also the QFT associated with the cyclic group $ \mathbb{Z} / N \mathbb{Z} $. In fact, any \textit{dephased} complex $ N \times N $ Hadamard matrix would do, but generally there are many other possible choices.
	
	We now describe the algorithm in detail, denoting the state after the $i$th step by $ \ket{\psi_i} $:
	\begin{enumerate}
		\setcounter{enumi}{-1}
		\item Prepare $ \ket{\psi_0} = \ket{0} $.
		
		\item Perform the unitary $V$, obtaining the state
		\begin{equation}
			\ket{\psi_1} = V \ket{0} = \frac{1}{\sqrt{N}} \sum_{x=0}^{N-1} \ket{x} .
		\end{equation}
		
		\item Query the oracle, and obtain:
		\begin{equation}
			\ket{\psi_2} = U_f \ket{\psi_1} = \frac{1}{\sqrt{N}} \sum_{x=0}^{N-1} \left( -1 \right)^{f \left( x \right)} \ket{x} .
		\end{equation}
		
		\item Act on the system using the unitary $ W = V^\dagger $, resulting in $ \ket{\psi_3} = V^\dagger \ket{\psi_2} $.
		
		\item Measure $ \ket{\psi_3}$ in the computational basis. Return ``constant'' iff we measured the state $ \ket{0} $, otherwise return ``balanced''.
	\end{enumerate}
	To see why the algorithm works, note:
	\begin{equation}\label{inner_product}
		\left\langle \psi_1 \vert \psi_2 \right\rangle = \frac{1}{N} \sum_{x=0}^{N-1} \left( -1 \right)^{f \left( x \right)} = \begin{cases}
			\pm 1 ; & \text{$f$ is constant} \\
			0 ; & \text{$f$ is balanced}
		\end{cases} .
	\end{equation}
	Thus, $ \ket{\psi_2} $ is either proportional to $ \ket{\psi_1} $ or orthogonal to it, based on whether $ f $ is constant or balanced. This means we obtain an orthogonal decomposition:
	\begin{equation}\label{decomp}
		\mathcal{H} = \mathcal{H}_c \oplus \mathcal{H}_b ,
	\end{equation}
	where $ \mathcal{H}_c = \mathrm{span}_{\mathbb{C}} \left\{ \ket{\psi_1} \right\} $ and $ \mathcal{H}_b = \mathcal{H}_c^\perp $. \eqref{inner_product} means that $ \ket{\psi_2} \in \mathcal{H}_j $ iff $f$ belongs to the class $j \in \left\{ c,b \right\}$, where $c$ and $b$ stand for ``constant'' and ``balanced'' respectively. To conclude the computation, we note that $V$ maps the standard basis to a basis adapted to the decomposition \eqref{decomp} (the first column of $V$ spans $ \mathcal{H}_c $, and the others are orthogonal to it, hence they span $ \mathcal{H}_b $). Therefore, applying $V^\dagger$ results in mapping $ \mathcal{H}_c $ to $ \ket{0} $ and $ \mathcal{H}_b $ to its orthogonal complement, so the algorithm indeed succeeds with probability $1$.
	
	From the above description of the DJ problem and algorithm, it is clear that no group structure on the set $ \left[ N \right] $ is assumed, and the algorithm does not utilize any such structure. In particular, the DJ algorithm does not necessarily perform a QFT. Instead, the promise on $f$ reflects a coarse structure, compared with the strict algebraic structure underlying the HSP (see \Cref{sec:HSP}). We can describe the promise on $f$ geometrically.
	Consider the computational-basis representation of the vector $ \ket{\psi_2} $. This column vector is proportional to $ \vec{g} \defeq \left( -1 \right)^{\vec{f}} $, where $ \vec{f} \in \mathbb{F}_2^N $ is the vector defined by $ f_x \defeq f \left( x \right) $, and the exponential is entry-wise. Actually it is even better to think of $ \vec{f} $ as a point in the affine space $ \mathbb{A}_{\mathbb{F}_2}^N $, but here we do not need this distinction. Geometrically, the task is to find out whether $ \vec{g} $ is proportional to $ \vec{1} $ (the all-ones vector) or perpendicular to it. A classical query can only ask about the values of $g_x = \left( -1 \right)^{f_{x}} $, the coordinates of $ \vec{g} $ with respect to the standard basis; whereas a quantum algorithm can also rotate $ \vec{g} $, thus solving the problem with a single query.
	
	The following subsection deals with straightforward generalizations of this idea. In \Cref{sec:idx2HSP} we endow the domain of $f$ with a group structure and refine the promise, to obtain a generalization of DJ that simultaneously serves as a special case of the HSP.
	
	\subsection{Generalizations of Deutsch--Jozsa}\label{sec:DJH}
	In the Deutsch--Jozsa--H{\o}yer problem~\cite{hoyer1999conjugated}, one is given a function $f : S \to G$ with $S$ a finite set and $G$ a set endowed with an addition operation; and the task is to decide whether $f$ is constant or not. If $G$ is an abelian group and we are promised that $f$ is either constant or balanced, then the problem can be solved with certainty using just a single query. Here ``balanced'' means that the preimages $f^{-1} \left( \left\{ g \right\} \right)$ have the same size for every $g \in G$ (we assume that $\abs{G}$ divides $\abs{S}$).
	
	The algorithm itself is also a direct generalization of DJ. Here we discuss the version where $G$ is an abelian group. Let $\chi$ be any nontrivial character of $G$, and use the trick from \Cref{sec:trick} to implement the phase oracle as defined in \eqref{phase_oracle}. As before, the algorithm performs the unitary $V$ on the initial state $\ket{0}$, applies the phase oracle, then applies $V^\dagger$ and measures in the computational basis. After the oracle query, we obtain the state:
	\begin{equation}
		\ket{\psi_2} \defeq U_f^{\mathrm{phase}} V \ket{0} = \frac{1}{\sqrt{\abs{S}}} \sum_{x \in S} \chi \left( f \left( x \right) \right) \ket{x} ,
	\end{equation}
	hence, the amplitude for measuring $0$ is:
	\begin{equation}\label{Hoyer_proof}
		\left\langle 0 \vert V^\dagger \vert \psi_2 \right\rangle = \frac{1}{\abs{S}} \sum_{x,y \in S} \left\langle y \vert \chi \left( f \left( x \right) \right) \vert x \right\rangle = \frac{1}{\abs{S}} \sum_{x \in S} \chi \left( f \left( x \right) \right) .
	\end{equation}
	If $f$ is constant, i.e. if there exists some $g \in G$ such that $f \left( x \right) = g$ for all $x \in S$, then we obtain $ \left\langle 0 \vert V^\dagger \vert \psi_2 \right\rangle = \chi \left( g \right) $, which is some complex number with modulus $1$;
	and as before, the probability to measure the state $\ket{0}$ at the end of the algorithm is one. However, if $f$ is balanced, then every $g \in G$ appears as the input for $\chi$ in \eqref{Hoyer_proof} exactly $ \abs{S} / \abs{G} $ times. Therefore:
	\begin{equation*}
		\frac{1}{\abs{S}} \sum_{x \in S} \chi \left( f \left( x \right) \right) = \frac{1}{\abs{G}} \sum_{g \in G} \chi \left( g \right) ,
	\end{equation*}
	which is zero by orthogonality to the trivial character. We can see that the group structure of $G$ plays a role here, but this role is relatively insignificant compared with the HSP.
	
	Further generalizations~\cite{batty2006extending,vicary2013topological} address the case of non-abelian $G$. However, to do so one should apply some non-abelian version of the ``shift-to-phase'' trick, which does not result in an actual shift oracle. These generalizations circumvent this issue by assuming somewhat non-intuitive promises on matrix elements of irreps of $G$.
	
	\section{The index-$q$ hidden subgroup problem}\label{sec:idx2HSP}
	In this section we consider a special case of the HSP, where $H$ is a normal subgroup promised to either be $G$ or have index $q$ in $G$ (so the index is either $1$ or $q$, for a known integer $q>1$), and $ \abs{X} = q $. We then describe a quantum algorithm that solves the problem using a single query, given certain conditions. In fact, our algorithm is a variant of both Deutsch--Jozsa--H{\o}yer (described in \Cref{sec:DJH}) and Shor--Kitaev algorithms. We first treat the case of abelian $G$, and in \Cref{sec:nonabelian} present the non-abelian version.
	
	We start with a heuristic explanation of the key idea.
	As in \Cref{sec:HSP}, let $f : G \rightarrow X$ denote a function that hides $H$. 
	The basic insight is that $f$ is either constant or balanced (in the sense of \Cref{sec:DJH}), depending on whether the index $ \left[ G : H \right]$ is $1$ or $q$.
	Hence, we can determine whether $H$ has index $1$ or $q$ using the generic Deutsch--Jozsa--H{\o}yer algorithm. Suppose $X$ has an arbitrary $\Zq$ structure, and we are given access to $f$ via the shift oracle. We convert the shift oracle to a phase oracle (as explained in \Cref{sec:trick}) with respect to the \emph{primitive} character $ \chi \left( x \right) = \omega^x $, where $ \omega \coloneqq e^{2\pi i/q} $ (we can replace $\omega$ by any other $q$th primitive root of unity). On the $G$ register, we perform the following procedure:
	\begin{enumerate}\setcounter{enumi}{-1}
		\item Prepare $ \ket{\psi_0} $ in a computational basis state.
		
		\item Apply any unitary $V$ such that $ \left\langle x \vert V \vert \psi_0 \right\rangle $ is the same for all computational basis states $ \ket{x} $.
		
		\item Perform a query using the phase oracle.
		
		\item Apply the unitary $V^\dagger$.
		
		\item Measure in the computational basis. If we measured $ \ket{\psi_0} $ then $f$ is constant, otherwise it is balanced.
	\end{enumerate}
	As we proved in \Cref{sec:DJH}, this algorithm always outputs the correct answer. 
	
	So far we made no use of the group structure on $G$. We now specify the initial state $ \ket{\psi_0} $ and unitary $V$, using this group structure. We use the Hilbert space $ \mathbb{C} G $; as explained in \Cref{sec:QFT}, we can identify the computational basis with the elements of either $G$ or $\hat{G}$. We start with $\hat{G}$, and with each (regular or inverse) QFT we toggle back and forth between the two perspectives.
	We take $ \ket{\psi_0} = \ket{\rho_0} $ (corresponding to the trivial character $ \rho_0 \in \hat{G} $),	and choose $V$ to be the inverse QFT with respect to $G$. We thus obtain an algorithm that combines the Deutsch--Jozsa--H{\o}yer and Shor--Kitaev algorithms; moreover, it	always discerns between the index-$1$ and index-$q$ cases. Thus, we are left with the following question: under which conditions does the final measurement determine $H$ completely?
	
	\hypertarget{alg:identification}{\textbf{The single-query identification algorithm.}}
	The group $G$ is known through a fixed presentation, with fixed labels for its elements and Fourier basis (in the Abelian case we use cyclic-factor coordinates). The response register is $X= \Zq$, with fixed known addition, and the oracle acts as $U_f^\mathrm{shift} \ket{g} \otimes \ket{x} = \ket{g} \otimes \ket{x + f(g) \bmod q}$. We convert it to the phase oracle $ U_f^\mathrm{phase} \ket{g} = \chi ( f (g) ) \ket{g} $ using the shift-to-phase trick, where $ \chi (x) = e^{2\pi i x /q} $.
	The algorithm's output is the measured character label $\rho$, representing the subgroup $ \tilde{H} \coloneqq \ker \rho$. We count only oracle calls; state preparation, Fourier transforms and classical processing are outside this cost measure.
	The algorithm is as follows:
	\begin{enumerate}\setcounter{enumi}{-1}
		\item Prepare $ \ket{ \psi_0 } = \ket{\rho_0} $, the computational basis state corresponding to the trivial character $ \rho_0 \in \hat{G} $.
		
		\item Perform an inverse QFT; obtain the democratic superposition
		\begin{equation}\label{eqn:alg1}
			\ket{\psi_1} = \ket{ \hat{\delta}_{\rho_0} } = \frac{1}{\sqrt{\abs{G}}} \sum_{g \in G} \ket{g} .
		\end{equation}
		
		\item Perform a single oracle query, obtaining the state:
		\begin{equation}\label{eqn:alg2}
			\ket{\psi_2} = \frac{1}{\sqrt{\abs{G}}} \sum_{g \in G} \chi ( f ( g ) ) \ket{g} .
		\end{equation}
		
		\item Perform the quantum Fourier transform of $G$, to obtain:
		\begin{equation}\label{Index_q_HSP_psi_3}
			\ket{\psi_3} = \ket{ \widehat{\chi \circ f} } = \sum_{\rho \in \hat{G}} \underbrace{\left( \frac{1}{\abs{G}} \sum_{g \in G} \chi ( f ( g ) ) \rho ( g ) \right)}_{\eqqcolon \hat{a}_\rho} \ket{\rho} ,
		\end{equation}
		where $ \widehat{\chi \circ f} $ denotes the Fourier transform of $ \chi \circ f $.
		
		\item Measure in the computational basis, now identified with the elements of $ \hat{G} $. Obtain a character $\rho$, and output the subgroup $\tilde{H} = \ker \rho \defeq \left\{ g \in G \mid \rho ( g ) = 1 \right\}$.
	\end{enumerate}
	
	We now seek sufficient conditions for the algorithm to succeed, i.e., for $ \tilde{H} = H $. We start by computing $\hat{a}_\rho$ (the parenthesized expression in \eqref{Index_q_HSP_psi_3}). As we have already explained, if $ \left[ G : H \right] = 1 $ then $f$ is constant and the final state must be $ \ket{\rho_0} $ (up to phase); and if $ \left[ G : H \right] = q $ then $f$ is balanced and the final state must be orthogonal to $ \ket{\rho_0} $. Thus, hereon we assume $ \left[ G : H \right] = q $.
	
	We proceed by choosing an arbitrary representative $ g_c \in G $ for each coset $c \in G/H$; we can write each group element uniquely as $ g = g_c \oplus h $. Thus, we split the sum in $ \hat{a}_\rho $ into two sums---one over all cosets, and the other over all elements in each coset. This yields:
	\begin{equation}\label{eqn:a_hat}
		\begin{split}
			\hat{a}_\rho &= \frac{1}{\abs{G}} \sum_{c \in G/H} \sum_{h \in H} \chi \left( f \left( g_c \oplus h \right) \right) \rho \left( g_c \oplus h \right) = \\
			&= \frac{\abs{H}}{\abs{G}} \left( \sum_{c \in G/H} \chi ( f ( g_c ) ) \rho ( g_c ) \right) \left( \frac{1}{\abs{H}} \sum_{h \in H} \rho ( h ) \right) ,
		\end{split}
	\end{equation}
	where we have used the facts that $f$ hides $H$ (hence only depends on the coset), and that $\rho$ is a homomorphism. Just as we saw in \Cref{sec:HSP}, the second parentheses equal the inner product of $ \rho \vert_H $ with the trivial character of $H$, which equal $1$ if $ \rho \in H^\perp $ and $0$ otherwise. Thus, just like in the Shor--Kitaev algorithm, we can only sample characters $\rho$ with $ H \le \ker \rho $; in other words, $ H \le \tilde{H} $ automatically, without any assumptions.
	
	Hereon, suppose $ \rho \in H^\perp $. We have the expression $ \hat{a}_\rho = \frac{\abs{H}}{\abs{G}} \sum_{c \in G/H} \chi ( f ( g_c ) ) \rho ( g_c ) $, which can be simplified as follows. Recalling that $ f = \iota \circ \pi $, we can write $ f ( g_c ) = \iota ( \pi ( g_c ) ) = \iota ( c ) $. Moreover, since $ \rho \in H^\perp $, it defines a character of $ G/H $; equivalently, it can be written in the form $ \rho = \lambda \circ \pi $ for $ \lambda \in \widehat{G/H} $. Thus, we can write $ \rho ( g_c ) = \lambda ( c ) $. Putting everything together:
	\begin{equation}
		\hat{a}_{\lambda \circ \pi} = \frac{\abs{H}}{\abs{G}} \sum_{c \in G/H} \chi ( \iota ( c ) ) \lambda ( c ) .
	\end{equation}
	This expression almost looks like an inner product between characters of $G/H$. However, $\iota : G/H \to X $ may not be a group homomorphism (since $f$ need not be a homomorphism), hence $ a \coloneqq \chi \circ \iota $ is not necessarily a character. Nevertheless, $ \hat{a}_{\lambda \circ \pi} $ is (up to normalization) the Fourier coefficient of the function $ a \in \mathbb{C} \left[ G/H \right] $ (corresponding to the character $ \lambda \in \widehat{G/H} $).
	
	Note that $ \tilde{H} = \ker \rho = \ker \left( \lambda \circ \pi \right) = \pi^{-1} \left( \ker \lambda \right) $. Thus, $ \tilde{H} = H $ if and only if $ \lambda $ is a faithful (injective) character. Since a non-cyclic group has no faithful characters, a necessary condition for our algorithm to output the correct answer is $ G / H \cong \Zq $ or $ G / H \cong 0 $ (the latter being the index-$1$ case). Moreover, if \emph{every} nontrivial character of $ G / H $ is faithful, then the algorithm is guaranteed to succeed. That is exactly the case for prime $q$; therefore, we deduce that for prime $q$ the algorithm identifies $H$ exactly with one query, without any further promises or assumptions.
	
	For composite (non-prime) $q$, the necessary and sufficient conditions are slightly more complicated. By the above discussion, the necessary and sufficient condition for identification of $H$ is that $ \hat{a}_{\lambda \circ \pi} = 0 $ for any non-faithful character $ \lambda \in \widehat{G/H} $. In other words, the vector space $ \mathbb{C} \left[ G / H \right] $ of complex-valued functions on $ G / H \cong \Zq $ has a basis corresponding to the characters $ \widehat{G/H} $; and the necessary and sufficient condition is that $ \chi \circ \iota $ belongs to the subspace spanned by the faithful characters.
	\begin{definition}\label{def:Fourier-support}
		We say that $\iota$ satisfies the \emph{Fourier-support condition}, if the Fourier transform of $ a = \chi \circ \iota \in \mathbb{C} \left[ G / H \right] $ is supported only on faithful characters of $G/H$.
	\end{definition}
	Our analysis above proves the following.
	\begin{theorem}\label{thm:identification_iff_Fourier_support}
		Let $G$ be a finite abelian group, $q>1$ an integer, $X \cong \Zq$, and $\chi \in \hat{X}$ the character $ \chi \left( x \right) = \omega^x $. Suppose the function $ f : G \to X $ hides a subgroup $H \le G$ such that either $H=G$ or $ G/H \cong \Zq $. Then, \hyperlink{alg:identification}{the single-query identification algorithm} outputs $\tilde{H} = H$ with certainty if and only if $H=G$ or the function $\iota$ satisfies the Fourier-support condition.
	\end{theorem}
	
	Next, we would like to make the Fourier-support condition more concrete. Note that the identification of $G/H$ with $\Zq$ need not agree with the $\Zq$-labels of $X$; in other words, the coset-to-label map $\iota$ is not necessarily the identity map, or any other group isomorphism, or even a homomorphism. Without additional information, we can only guarantee that $\iota$ is a bijection of sets. Let us illustrate the problem with an example: suppose $ G/H \cong X \cong \Zfour $ and $ \chi ( x ) = i^x $. The characters of $ \Zfour $ are $ 1 $ (the trivial character), $ \chi $, $ \chi^2 $ and $ \chi^3 $. Note that $ \lambda_2 \coloneqq \chi^2 $ is not faithful, since $ \chi^2 \left( 2 \right) = \left( -1 \right)^2 = 1 $; so we would like $ \hat{a}_{\lambda_2 \circ \pi} $ to vanish. If $\iota$ is a group isomorphism, then $ \chi \circ \iota $ is a faithful character; hence it must be either $ \chi $ or $ \chi^3 $ (depending on whether $ \iota $ is the identity map or the nontrivial $\Zfour$-automorphism, multiplication by $3$ modulo $4$). However, $\iota$ can be an arbitrary permutation in $\mathrm{Sym} \left( \Zq \right) $, e.g. $\iota = \left( 0 1 \right)$, in which case:
	\begin{equation}
		\begin{split}
			\hat{a}_{\lambda_2 \circ \pi} = \frac{1}{4} \sum_{c=0}^3 \chi(\iota(c)) \lambda_2(c) = \frac{1}{4} \sum_{c=0}^{3} i^{\iota(c)} (-1)^c = \frac{1}{4} \left( i^{\iota(0)} - i^{\iota(1)} + i^{\iota(2)} - i^{\iota(3)} \right) = \\
			= \frac{1}{4} \left( i^1 -i^0 +i^2 -i^3 \right) = \frac{1}{4} \left( i -1 -1 +i  \right) = \frac{-1+i}{2} ,
		\end{split}
	\end{equation}
	hence we have a probability of $ \abs{(i-1)/2}^2 = 1/2 $ to output $ \tilde{H} = \ker \left( \lambda_2 \circ \pi \right) = \pi^{-1} \left( \ker \lambda_2 \right) = \pi^{-1} \left( \left\{ 0, 2 \right\} \right) $, which is the subgroup of $G$ comprising the union of $H$ with the $H$-coset that corresponds to $2$ in the quotient $G/H$.
	
	More generally, if $q$ is composite, there exists an integer $ 1 < a < q $ such that $ a \vert q $; and the $\Zq$-character $ \lambda_a \left( c \right) \coloneqq \omega^{ac} $ is always nontrivial and non-faithful (e.g. $ \lambda_a \left( q/a \right) = 1 $). We have:
	\begin{equation}
		\hat{a}_{\lambda_a \circ \pi} = \frac{1}{q} \sum_{c=0}^{q-1} \chi(\iota(c)) \lambda_a (c) = \frac{1}{q} \sum_{c=0}^{q-1} \omega^{\iota \left( c \right)} \omega^{ac} .
	\end{equation}
	We know that the sum vanishes for $ \iota = \mathrm{id} $. The swap $ \iota = \left( 0 1 \right) $ changes only the first two terms, so it suffices to show that
	\begin{equation}
		\begin{split}
			& \omega^{\iota \left( 0 \right)} \cdot \omega^{a \cdot 0} + \omega^{\iota \left( 1 \right)} \cdot \omega^{a \cdot 1} \ne 
			\omega^{0} \cdot \omega^{a \cdot 0} + \omega^{1} \cdot \omega^{a \cdot 1} \\
			\Leftrightarrow \quad & \omega + \omega^a \ne 1+ \omega^{a+1} \\
			\Leftrightarrow \quad & \omega -1 \ne \omega^{a} \left( \omega -1 \right) \\
			\Leftrightarrow \quad & \omega \ne 1 \; \wedge \; \omega^a \ne 1
		\end{split} ,
	\end{equation}
	which is obvious since $ a < q $. Therefore, for any composite $q$ the swap $ \iota = \left( 0 1 \right) $ fails the algorithm.
	
	To characterize the full set of valid bijections $\iota$ for arbitrary $q$, one can reformulate the Fourier-support condition geometrically. A character $\lambda \in \widehat{G/H}$ is non-faithful if and only if it is trivial on at least one minimal non-trivial subgroup of $\mathbb{Z}/q\mathbb{Z}$. These minimal subgroups have prime order $p$ (generated by the element $q/p$) for each prime factor $p$ of $q$. Standard Fourier inversion over finite abelian groups yields a necessary and sufficient condition for the vanishing of the the amplitudes $\hat{a}_{\lambda \circ \pi}$ for all non-faithful characters:
	\begin{lemma}
		Let $A$ be a finite abelian group, and let $ a \in \mathbb{C} \left[ A \right] $.
		The Fourier coefficient $ \hat{a} ( \lambda ) = \frac{1}{\abs{A}} \sum_{c \in A} a ( c ) \lambda ( c ) $ vanishes for all non-faithful characters $ \lambda \in \hat{A} $ if and only if the sum $ \sum_{i \in I} a ( d+i ) $ vanishes for every nontrivial minimal subgroup $ I \le A $ and every coset $d+I$.
	\end{lemma}
	\begin{proof}
		Let $ \lambda \in \widehat{A} $ be a non-faithful character; then $ \ker \lambda $ contains at least one minimal non-trivial subgroup $I$.
		We choose a representative from each coset $ d \in A / I $, and rewrite $ \hat{a} ( \lambda ) $ by splitting the sum into a double sum:
		\begin{equation}
			\begin{split}
				\hat{a} (\lambda) = \frac{1}{\abs{A}} \sum_{d \in A /I} \sum_{i \in I} a(d+i) \lambda(d+i) = \frac{1}{\abs{A}} \sum_{d \in A /I} \sum_{i \in I} a(d+i) \lambda ( d ) \underbrace{\lambda ( i )}_{=1} = \\
				= \frac{1}{\abs{A}} \sum_{d \in A /I} \lambda ( d ) \left( \sum_{i \in I} a ( d+i ) \right)
			\end{split} .
		\end{equation}
		Therefore, if the sum $ \sum_{i \in I} a ( d+i ) $ vanishes for every such $I$ and $d$, then $ \hat{a} (\lambda) $ vanishes.\\
		Conversely, suppose $ \hat{a} ( \lambda ) = 0 $ for every non-faithful $\lambda$, and let $ I \le A $ be a minimal non-trivial subgroup. Then, using the Fourier inversion formula $ a ( c ) = \sum_{\lambda \in \widehat{A}} \hat{a} ( \lambda ) \lambda^* ( c ) $, we find
		\begin{equation}
			\sum_{i \in I} a ( d+i ) = \sum_{i \in I} \sum_{\lambda \in \widehat{A}} \hat{a} ( \lambda ) \lambda^* ( d+i ) =
			\sum_{\lambda \in \widehat{A}} \hat{a} ( \lambda ) \lambda^* ( d ) \left( \sum_{i \in I} \lambda^* ( i ) \right) .
		\end{equation}
		The parentheses equal (up to normalization) the inner product between $ \lambda \vert_I $ and the trivial character of $I$. If $ \lambda $ is faithful then $ \lambda \vert_I $ is not trivial, hence the parentheses vanish; but if $ \lambda $ is not faithful, then by assumption $ \hat{a} ( \lambda ) = 0 $. Thus, every term in the outer sum is zero, which concludes the proof.
	\end{proof}
	Applying the lemma for $ A = G/H $ and $ a ( c ) = \omega^{ \iota ( c ) } $, we see that $\iota$ satisfies the Fourier-support condition if and only if the function $\omega^{\iota(c)}$ sums to zero over every coset of every nonzero minimal subgroup of $ \Zq $.
	Geometrically, this imposes a ``coset balance'' condition: for every prime factor $p$ of $q$, if we sample $\iota$ along any arithmetic progression of step $q/p$, the resulting roots of unity must form a perfectly balanced $p$-gon in the complex plane. For example, any \emph{affine relabeling} $ \iota ( c ) = \alpha c +\beta $ (where $ \alpha, \beta \in \Zq $ such that $ \mathrm{gcd} \left( \alpha, q \right) =1 $) naturally satisfies this property, as it maps arithmetic progressions to arithmetic progressions. However, non-affine permutations may also satisfy the coset-balance condition. For instance, in $\mathbb{Z}/8\mathbb{Z}$, the only minimal subgroup has order $2$ (step size $4$), so the sole requirement is $\omega^{\iota(c)} + \omega^{\iota(c+4)} = 0$, which simplifies to $\iota(c+4) \equiv \iota(c) + 4 \pmod 8$. One can easily construct non-affine permutations satisfying this local pairing (e.g., by swapping independent pairs), thereby yielding valid success-guaranteed bijections beyond affine relabelings.
	
	The following definition, proposition and corollary summarize the analysis above. 
	\begin{definition}\label{def:coset_balanced}
		A bijection of sets $ \iota : \Zq \to \Zq$ is \emph{coset-balanced} if for every prime factor $p$ of $q$, the sum $ \sum_{n \in A} \omega^{\iota ( n )} $ vanishes for all arithmetic progressions $A \subset \Zq$ with step size $q/p$.
	\end{definition}
	\begin{proposition}
		A bijection $ \iota : \Zq \to \Zq$ satisfies the Fourier-support condition if and only if it is coset-balanced.
	\end{proposition}
	\begin{corollary}
		Let $G$ be a finite abelian group, $q>1$ an integer, $X \cong \Zq$, and $\chi \in \hat{X}$ the character $ \chi ( x ) = \omega^x $. Suppose the function $ f : G \to X $ hides a subgroup $H \le G$ such that either $H=G$ or $ G/H \cong \Zq $. Then, \hyperlink{alg:identification}{the single-query identification algorithm} outputs $\tilde{H} = H$ with certainty if and only if $H=G$ or the function $\iota$ is coset-balanced.
	\end{corollary}
	
	This single-query exact identification represents a quantum advantage over classical methods.
	For $G = \left( \Zq \right)^n$ with $q$ prime and $n\ge2$, there are $M=(q^n-1)/(q-1)$ candidate index-$q$ subgroups. Fixing one valid hiding oracle per subgroup, each query returns one of $q$ possible labels. Thus, after $t$ queries, the oracle can produce at most $q^t$ distinct sequences of responses. Since a deterministic classical algorithm bases its final output entirely on this sequence, it can uniquely identify at most $q^t$ subgroups. Exact identification therefore requires $ t \ge \lceil \log_q M \rceil = n$. 
	Furthermore, by Yao's minimax principle, relaxing the requirement to allow a bounded error probability $\epsilon$ means the algorithm must correctly identify at least a $(1 - \epsilon)$ fraction of the $M$ uniform candidates. This strictly requires $q^t \ge (1 - \epsilon)M$, forcing any randomized classical algorithm to use $t \ge \log_q \big( (1 - \epsilon) M \big) = \Omega(n)$ queries. In contrast, our quantum algorithm achieves exact identification in a single query.
	This advantage generalizes the $\mathcal{O}(n)$ versus $\mathcal{O}(1)$ query separation of the Bernstein--Vazirani algorithm, which is a special case of our algorithm (see \Cref{sec:BV}). 
	
	Of course, as the minimal-label index-$q$ HSP is an instance of the finite abelian HSP, it can also be approached using the Shor--Kitaev algorithm. Recall (\Cref{sec:HSP}) that in each iteration, the Shor--Kitaev algorithm applies the QFT and samples a character $ \rho \in H^\perp $, where $ H^\perp \le \hat{G} $ is the subgroup of characters trivial on $H$. This subgroup is canonically isomorphic to $ \widehat{G/H} $: a character of $G$ defines a character of $G/H$ if and only if it factors via $\pi$, which occurs if and only if it is trivial on $H$. In our setting, $H^\perp \cong \widehat{G/H} \cong G/H $ has either $1$ or $q$ elements, depending on whether $[G:H]=1$ or $q$. Each element of $ H^\perp$ is sampled with equal probability, and from a sample $\rho$ one can compute $ \ker \rho $, which is a supergroup of $H$ (i.e. $H \le \ker \rho$). Exact recovery of $H$ therefore requires multiple samples and intersecting their kernels. For a \textit{single} sample, however, there are two limitations.
	First, there is always a nonzero probability of obtaining the trivial character $\rho_0$, which yields no information. Second, even if $\rho \neq \rho_0$, we only identify $H$ exactly when $\rho$ is faithful on $G/H$, which occurs iff $G/H$ is cyclic and $\rho = \omega^m$ with $\gcd \left( m, q \right) = 1$.
	Consequently, in the non-cyclic case a single Shor--Kitaev sample does not allow exact identification of $H$ with certainty, while in the cyclic case the best-case success probability is $\varphi(q)/q$ (equal to $1-1/q$ when $q$ is prime). In contrast, our algorithm (under the stated assumptions) always produces a character $\rho$ with $\ker\rho=H$ in a single query.
	
	\subsection{Example: $q=2$}\label{sec:index_2_HSP}
	Let $X \cong \Ztwo $ structure, and as in \Cref{sec:DJ} we use the alternating irrep $\chi$ to convert the shift oracle for $f$ into a phase oracle. The map $ \iota : G/H  \to X$ is either a constant function from a singleton to a two-element set, or a bijection of two-element sets. In both cases it must be either a group homomorphism, or a group homomorphism post-composed with a swap of the two elements of $X$. Therefore, $ \chi \circ f : G \rightarrow \left\{ \pm 1 \right\} $ is a group homomorphism, perhaps up to a minus sign. In fact, for all $g \in G$:
	\begin{equation}\label{chi_f_is_rep}
		\chi \circ f \left( g \right) = \pm \chi \circ \pi \left( g \right) = \pm \left( -1 \right)^{\pi \left( g \right)} = \pm \rho_H \left( g \right) ,
	\end{equation}
	where $ \rho_H \left( g \right) \defeq \left( -1 \right)^{\pi \left( g \right)} $ is defined by identifying the elements of $G / H$ with $0$ and $1$ in the obvious way if $H$ has index $2$; and if $H=G$, $\rho_H$ is defined to be the trivial representation of $G$. In the former case, $\rho_H$ is simply a representation of $G$ which is $1$ on $H$ and $-1$ on the other coset.
	
	Thus, the algorithm we described above yields $ \ket{\psi_3} = \ket{ \rho_H } $ (up to a global phase). We measure $ \rho_H $ with probability $1$, and $ \ker \rho_H = H $, as required.
	Note we have $ \rho_H = \rho_0 $ (the trivial character) iff the function $f$ is constant; but with a single query we obtain more information than that. We know whether $H$ has index $1$ or $2$; and if it has index $2$ we know which subgroup it is. This is somewhat analogous to how Simon's algorithm solves two versions of Simon's problem: the decision-problem-version and the function-problem-version.
	
	To illustrate the minimal-label index-$2$ HSP with an example, we consider a refinement of the Deutsch--Jozsa problem with $n=2$. Suppose we are given a function $ f : \left\{ 0,1 \right\}^2 \rightarrow \left\{ 0,1 \right\} $ that is promised to be either constant or balanced. If we identify $ \left\{ 0,1 \right\}^2 $ with the group $G = \left( \Ztwo \right) \times \left( \Ztwo \right) $ (the Klein four-group), $f$ hides a subgroup $H$. $H$ is either $G$ (if $f$ is constant) or an index-$2$ subgroup (if $f$ is balanced).
	
	Let $ \mathcal{F} $ be the set of admissible functions, i.e. functions $f$ that get two bits as input and output one bit, and are either constant or balanced. Let $\mathrm{left}$ ($\mathrm{right}$) be the function that outputs its left (right) bit; $ \mathrm{xor} $ outputs the exclusive OR of its two bits; and the prefix $\mathrm{n}$ indicates that the function is followed by a NOT. For example, $ \mathrm{nleft} \left( 01 \right) = 1 $, since the function takes the left bit ($0$) and then applies a NOT, flipping its value. The set $ \mathcal{F} $ is partitioned as follows:
	\begin{equation}\label{Klein_partition}
		\mathcal{F} = \left\{ f_0, f_1 \right\} \cup \left\{ \mathrm{left}, \mathrm{nleft} \right\} \cup \left\{ \mathrm{right}, \mathrm{nright} \right\} \cup \left\{ \mathrm{xor}, \mathrm{nxor} \right\},
	\end{equation}
	based on whether $f$ hides $G$, $\left\langle \left( 0,1 \right) \right\rangle $, $\left\langle \left( 1,0 \right) \right\rangle$ or $\left\langle \left( 1,1 \right) \right\rangle$ respectively (where $\left\langle g \right\rangle$ denotes the subgroup of $G$ generated by $g$). 
	The generic algorithm described above reduces to the Deutsch--Jozsa algorithm, and it determines to which element of the partition \eqref{Klein_partition} $f$ belongs.
	
	The obstacle for generalizing this example directly for higher values of $n$, is that for $n>2$ not every partition of $ \left( \mathbb{Z} / 2 \mathbb{Z} \right)^n $ to two equally-sized subsets corresponds to cosets of some subgroup. For example, with $n=3$ we have the following subset of size $4$: $ \left\{ 000, 001, 010, 100 \right\} $, which contains the identity but is not a subgroup.
	
	\subsection{Non-abelian version}\label{sec:nonabelian}
	In this subsection, we generalize the minimal-label index-$q$ hidden subgroup problem, such that the ambient group $G$ need not be abelian (but is still assumed to be finite). Thus, we explicitly assume that the hidden subgroup $H \le G$ is normal; and the algorithm uses the non-abelian version of the quantum Fourier transform. Remarkably, with these slightly modified assumptions, we get essentially the same result: the single-query identification algorithm distinguishes between index $1$ and $q$ without any further assumptions; and for index $q$ and cyclic quotient, it correctly identifies the hidden subgroup if $\iota$ satisfies the Fourier-support condition. This upshot is summarized in \Cref{thm:non_abelian}. We start by presenting the non-abelian version of the quantum Fourier transform; discuss the necessary modifications to the algorithm compared with the abelian case; and then analyze the algorithm's performance and derive \Cref{thm:non_abelian}.
	
	Let $G$ be a finite non-abelian group, and let $\hat{G}$ denote the equivalence classes of finite-dimensional irreducible representations of $G$. Since $G$ is non-abelian, its irreps need not be one-dimensional; hence the elements of $\hat{G}$ cannot be interpreted as functions $ G \to \mathbb{C} $, and in particular do not form a group. However, the left regular representation $ \CG $ decomposes as follows:
	\begin{equation}\label{eqn:non_abelian_FT}
		F_G : \CG \xrightarrow{\sim} \bigoplus_{\rho \in \hat{G}} \End \left( V_\rho \right) ,
	\end{equation}
	where $V_\rho$ is a vector space carrying the irrep $\rho$, and $G$ acts on $ \End \left( V_\rho \right) $ via post-composition with $ \rho (g) $. Thus, $ \End \left( V_\rho \right) $ is isomorphic (non-canonically) to $ d_\rho \coloneqq \dim V_\rho $ copies of $ V_\rho $. As in the case of abelian $G$, the Fourier transform $F_G$ implements the isomorphism \eqref{eqn:non_abelian_FT}, defined as follows:
	\begin{equation}
		F_G \left( f \right) = \bigoplus_{\rho \in \hat{G}} \sqrt{ \frac{d_\rho}{ \abs{G} } } \sum_{g \in G} f(g) \rho (g) ,
	\end{equation}
	where $ f \in \CG $ is interpreted as a complex function on $G$: writing $f = \sum_{g \in G} f_g g $ as a formal sum of $G$ elements, $f(g)$ is the complex coefficient $f_g$. Recall the left regular representation acts on $f$ as
	\begin{equation}
		\forall h \in G, \quad h \cdot f = \sum_{g \in G} f_g hg = \sum_{g' \in G} f_{h^{-1} g'} g' ,
	\end{equation}
	so $ \left( h \cdot f \right) (g) = f ( h^{-1}g ) $, and we obtain
	\begin{equation}\label{eqn:non_abelian_shift_phase}
		\begin{split}
			F_G ( h \cdot f ) = \bigoplus_{\rho \in \hat{G}} \sqrt{ \frac{d_\rho}{ \abs{G} } } \sum_{g \in G} f( h^{-1} g) \rho (g) = \bigoplus_{\rho \in \hat{G}} \sqrt{ \frac{d_\rho}{ \abs{G} } } \sum_{g' \in G} f( g' ) \rho (hg') = \\
			= \bigoplus_{\rho \in \hat{G}} \rho \left( h \right) \sqrt{ \frac{d_\rho}{ \abs{G} } } \sum_{g' \in G} f( g' ) \rho (g') = D_h F_G (f)
		\end{split} ,
	\end{equation}
	as required, where $D_h$ is the $G$-representation on $ \bigoplus_{\rho \in \hat{G}} \End \left( V_\rho \right) $ mentioned above, acting on each summand as post-composition by $ \rho (h) $. \eqref{eqn:non_abelian_shift_phase} is the non-abelian version of the ``shift-to-phase'' property \eqref{phase_to_shift}. The inverse Fourier transform is given by
	\begin{equation}
		F_G^{-1} \left( \bigoplus_{\rho \in \hat{G}} A_\rho \right) (g) = \frac{1}{\sqrt{\abs{G}}} \sum_{\rho \in \hat{G}} \sqrt{ d_\rho } \tr{ A_\rho \rho ( g^{-1} ) } .
	\end{equation}

	To get a quantum version of this Fourier transform, we need to treat both spaces in \eqref{eqn:non_abelian_FT} as Hilbert spaces carrying \emph{unitary} representations of $G$. We do this for $\CG$ as in the abelian case, i.e., identify $ \CG \cong \spanC \left\{ \ket{g} \right\}_{g \in G} $. For the right hand side of \eqref{eqn:non_abelian_FT}, we choose an inner product for each $ V_\rho $ such that $\rho$ is unitary (this can always be done via the ``averaging trick''), and then take the Frobenius inner product on $ \End \left( V_\rho \right) $. Under these conventions, the transformation $F_G$ is unitary. Note that unlike the abelian case, the Hilbert space $ \bigoplus_{\rho \in \hat{G}} \End \left( V_\rho \right) $ does not have a distinguished orthonormal basis; consequently, it is more convenient to describe an entire linear map $A_\rho \in \End \left( V_\rho \right) $ as a single cohesive unit, rather than choosing a basis for $V_\rho$ and keeping track of individual matrix elements.
	
	We can now describe the non-abelian version of \hyperlink{alg:identification}{the single-query identification algorithm}. The assumptions are precisely the same as in the abelian case, except that $G$ need not be abelian; in particular, $H \le G$ is assumed to be a \emph{normal} subgroup, such that $ G/H \cong \Zq $ or $H=G$. The identification algorithm prepares the state $ \ket{\psi_0} = \ket{\rho_0} $, which corresponds to the element of $ \bigoplus_{\rho \in \hat{G}} \End (V_\rho) $ with $1$ on the summand $ \End (V_{\rho_0}) \cong \End(\mathbb{C}) $ corresponding to the trivial representation and $0$ elsewhere. We then apply the non-abelian inverse quantum Fourier transform, which results in the same state $ \ket{\psi_1} $ as in \eqref{eqn:alg1}, followed by an oracle query yielding $ \ket{\psi_2} $ from \eqref{eqn:alg2}. The next step is the non-abelian Fourier transform, which yields
	\begin{equation}
		\ket{\psi_3} = \sum_{\rho \in \hat{G}} \underbrace{\left( \frac{ \sqrt{d_\rho} }{\abs{G}} \sum_{g \in G} \chi ( f ( g ) ) \rho ( g ) \right)}_{\eqqcolon \hat{a}_\rho} \ket{\rho} ,
	\end{equation}
	where $ \rho (g) \ket{\rho} $ is shorthand for $ \sum_{i,j=1}^{d_\rho} \left[ \rho (g) \right]_{ij} \ket{\rho; i,j} $, for $i,j$ indices of an orthonormal basis for $V_\rho$ chosen as explained above. However, as we mentioned, it is convenient to keep track of the entire linear transformation $ \hat{a}_\rho = \frac{ \sqrt{d_\rho} }{\abs{G}} \sum_{g \in G} \chi (f(g)) \rho(g) $, rather than individual matrix coefficients. Ostensibly, the final measurement produces not only an irrep $ \rho \in \hat{G} $ but also a pair of indices $i,j$; one should ignore the indices and output $ \hat{H} = \ker \rho $ as in the abelian case. As we shall see, the probability of measuring an irrep $\rho$ with $d_\rho >1$ is zero, so the instruction to ignore the indices is redundant.
	
	The algorithm's analysis is very similar to the abelian case. In what follows, we sketch the main differences. Again, if $H=G$ then $f$ is constant, and by Deutsch--Jozsa--H{\o}yer we get the trivial representation with probability $1$; so hereon we assume $ G/H \cong \Zq $.
	Following the same logic leading to \eqref{eqn:a_hat}, we have
	\begin{equation}
		\hat{a}_\rho = \sqrt{d_\rho} \frac{\abs{H}}{\abs{G}} \left( \sum_{c \in G/H} \chi ( f ( g_c ) ) \rho ( g_c ) \right) \underbrace{\left( \frac{1}{\abs{H}} \sum_{h \in H} \rho ( h ) \right)}_{P_\rho} 
	\end{equation}
	for arbitrarily-chosen representatives $ g_c $ of the cosets $ c \in G/H $. Note each parenthesis is a linear transformation on $V_\rho$; the right one, denoted $P_\rho$, is the projection onto the subspace $ V_\rho^H = \left\{ v \in V_\rho \mid \rho(h)v = v \text{ for all } h \in H \right\} $. In the abelian case, we had argued that this expression vanishes if $\rho \vert_H$ is nontrivial; we now do the same for the non-abelian case.
	
	First, we show that $V_\rho^H$ is $G$-invariant. Let $ v \in V_\rho^H $, $g \in G$, $h \in H$, and we wish to show that $ \rho(g)v $ is fixed by $ \rho(h) $, i.e., that $ \rho(h) \rho(g) v = \rho(g) v $. The LHS equals $ \rho(hg) $; and since $H$ is normal, there exists some $h' \in H$ such that $ hg=gh' $. Thus,
	\begin{equation}
		\rho(h) \rho(g) v= \rho(hg)v = \rho(gh')v = \rho(g) \rho(h') v = \rho(g) v
	\end{equation}
	as required. Now, since $V_\rho^H$ is a $G$-invariant subspace of the irreducible representation $V_\rho$, it must be either $ V_\rho $ or zero. If $ V_\rho^H = V_\rho $ then every $ v \in V_\rho $ is fixed by $H$, so $ \rho_H $ is the trivial representation and $ P_\rho $ is the identity matrix; and if $ V_\rho^H = 0 $ then $ P_\rho = 0 $.
	
	Thus, hereon we can assume that $ \rho \vert_H $ is trivial; in other words, $ H \le \ker \rho = \tilde{H} $. As we argued in the abelian case, this means that $\rho$ factors through the quotient $G/H$, i.e., $ \rho = \lambda \circ \pi $ for some representation $ \lambda $ of $G/H \cong \Zq $. Now, since $\rho$ is irreducible, $\lambda$ must also be irreducible; but a representation of an abelian group (such as $\Zq$) is irreducible if and only if it is one-dimensional. Therefore, the only non-zero coefficients $\hat{a}_\rho$ appear for one-dimensional irreps $\rho$ with $ \rho \vert_H $ trivial. Thus, everything reduces completely to the setup detailed in the analysis of the abelian case, and we obtain a non-abelian version of \Cref{thm:identification_iff_Fourier_support}:
	\begin{theorem}\label{thm:non_abelian}
		Let $G$ be a finite group, $q>1$ an integer, $X = \Zq$, and $\chi \in \hat{X}$ the character $ \chi \left( x \right) = \omega^x $. Suppose the function $ f : G \to X $ hides a normal subgroup $H \trianglelefteq G$ such that either $H=G$ or $ G/H \cong \Zq $. Then, the single-query identification algorithm outputs $\tilde{H} = H$ with certainty if and only if $H=G$ or the function $\iota$ satisfies the Fourier-support condition.
	\end{theorem}
	Of course, the Fourier-support condition can be replaced by the coset-balance condition.
	\begin{remark}\label{rmk:abelianization}
		In the above theorem, the promise on $H$ guarantees that $G/H$ is abelian. By the universal property of abelianization, the quotient map $ \pi : G \to G / H $ factors (uniquely) through the abelianization map, $ G \to G_\mathrm{ab} \coloneqq G / [G, G] $ (equivalently, the commutator subgroup $ [G, G] $ is contained in $ H $).
		Since the oracle is constant on cosets of $H$---and hence on cosets of $[G,G]$---the original problem of identifying $H \le G $ reduces to its ``abelianized'' version, namely finding the hidden subgroup $ H / [G, G] \le G_\mathrm{ab} $ (this reduction is similar to the one detailed in \Cref{sec:reduction}).
		The latter problem is solved by our algorithm using the abelian Fourier transform over $ G_\mathrm{ab} $; thus, a non-abelian Fourier transform is not strictly necessary.
	\end{remark}
	
	\section{The Bernstein--Vazirani algorithm}\label{sec:BV}
	The Bernstein--Vazirani (BV) problem and algorithm are very similar to Deutsch--Jozsa, at least superficially. Indeed, in both cases the oracle implements a function $ f : \left\{ 0,1 \right\}^n \rightarrow \left\{ 0, 1 \right\} $. Moreover, the BV algorithm uses the exact same unitary gates as the standard Deutsch--Jozsa algorithm---namely, a Hadamard transform before and after the single oracle query.
	However, as shown in \Cref{sec:DJ}, there exist infinitely many variants of the Deutsch--Jozsa algorithm that use different unitary gates.
	This begs the question, whether the BV algorithm could also be ``deformed'' in an essential way. The answer is no. This is because the BV problem \textit{does} use a particular group structure on $\left\{ 0,1 \right\}^n$, namely $ \left( \mathbb{Z} / 2 \mathbb{Z} \right)^n $.
	Thus, the BV algorithm directly utilizes the quantum Fourier transform of $ \left( \mathbb{Z} / 2 \mathbb{Z} \right)^n $, i.e. the Hadamard transform. As we shall see, the BV problem is an instance of the minimal-label index-$2$ HSP; and the BV algorithm is an instance of our algorithm from \Cref{sec:idx2HSP}.
	
	In the BV problem, we are tasked with identifying a function $f : \left( \mathbb{Z} / 2 \mathbb{Z} \right)^n \rightarrow \mathbb{Z} / 2 \mathbb{Z}$, promised to be of the form $ f \left( \vec{x} \right) = \vec{x} \cdot \vec{s} $ for some unknown vector $ \vec{s} $. Alternatively, $f$ is promised to be a linear map of vector spaces over the field $ \mathbb{F}_2 $; these two promises are equivalent since a linear map over $ \mathbb{F}_2 $ is the same thing as a group homomorphism (a linear map is by definition additive, i.e. a homomorphism of the abelian groups underlying the vector spaces; and the condition of respecting the scalars is vacuous since the scalars are only $0,1$). Since $f$ is a group homomorphism, it factors through the surjection $ \pi : \left( \mathbb{Z} / 2 \mathbb{Z} \right)^n \twoheadrightarrow \mathrm{im} f \cong \left( \mathbb{Z} / 2 \mathbb{Z} \right)^n / \ker f $. Thus, $f$ hides the subgroup $ H \defeq \ker f \le \left( \mathbb{Z} / 2 \mathbb{Z} \right)^n $, which is $ \left( \mathbb{Z} / 2 \mathbb{Z} \right)^n$ if $\vec{s}=0$ and has index $2$ otherwise. Conversely, for each index-$2$ (or index-$1$) subgroup $H \le \left( \mathbb{Z} / 2 \mathbb{Z} \right)^n $ there exists a unique homomorphism $f : \left( \mathbb{Z} / 2 \mathbb{Z} \right)^n \rightarrow \mathbb{Z} / 2 \mathbb{Z} $ with $ \ker f = H $: the one that maps $\vec{x}$ to $0$ if $ \vec{x} \in H $, and to $1$ otherwise. Thus, identifying the homomorphism $f$ is the same as identifying its kernel; so the BV problem is a special case of the minimal-label index-$2$ HSP.
	
	Since $2$ is prime, $\iota$ satisfies the Fourier-support condition automatically (see \Cref{def:Fourier-support}); in fact, since $f$ is a group homomorphism, so is $\iota$. Thus, the BV problem can be solved with a single query using our algorithm. While this description of the BV algorithm may differ from the standard one, the algorithm itself is identical.
	Recall from \Cref{sec:idx2HSP} that single-sample Shor--Kitaev for index $q=2$ succeeds with probability $ \varphi \left( 2 \right)/2 = 1/2$; in contrast, the BV circuit leverages the linear phase $(-1)^{\vec{x} \cdot \vec{s}} $ to recover the defining character (and hence $H$) with certainty in one query.
	
	\subsection{Reduction of the minimal-label index-$2$ hidden subgroup problem to Bernstein--Vazirani}\label{sec:reduction}
	In this subsection, we show that any instance of the minimal-label index-$2$ HSP with finite abelian $G$ reduces to the Bernstein--Vazirani problem. The idea is that elements of the form $ g + g \in G $ automatically belong to $H$, so we can quotient them all out without losing any pertinent information; and the resulting quotient is isomorphic to $ \left( \mathbb{Z} / 2 \mathbb{Z} \right)^n $ for some $n$.
	
	Let us elaborate. First, define the subgroup $ 2 G \coloneqq \left\{ g + g : g \in G \right\} \le G $, and let $ i : 2G \hookrightarrow G $ be the natural inclusion. The cokernel of $i$ is the projection $ p : G \twoheadrightarrow G / 2G $. By the universal property of cokernels, any homomorphism $ h : G \to A $ such that $ h \circ i = 0 $ factors uniquely through $p$; and this assumption holds for $\pi$, since $ \pi \left( g + g \right) = \pi \left( g \right) + \pi \left( g \right) $, which is zero because the image of $\pi$ is either $ \mathbb{Z} / 2 \mathbb{Z} $ or the trivial group. Thus, there exists a unique homomorphism $ \bar{\pi} : G / 2G \to G / H $ such that $ \pi = \bar{\pi} \circ p $.
	This means that the elements of $2G$ are completely inconsequential: the problem of finding the subgroup $ H \le G $ hidden by $f$ is equivalent to the problem of finding the subgroup $ H / 2G \le G / 2G $ hidden by $ \bar{f} \coloneqq \iota \circ \bar{\pi} : G/2G \to X $.
	Now, $ G / 2G $ is a finite abelian group where every nonzero element has order $2$, so it is isomorphic to $ \left( \mathbb{Z} / 2 \mathbb{Z} \right)^n $ for some $n$. Thus, $ \mathbb{C} \left[ G / 2G \right] $ is a Hilbert space of $n$ qubits, which can be implemented as a subspace of $ \mathbb{C} \left[ G \right] $ via the identification $ \ket{g + 2G} \coloneqq \frac{1}{\sqrt{ \abs{2G} }} \sum_{h \in 2G} \ket{ g+h } $ (i.e., we require that all basis elements belonging to the same $2G$ coset have the same coefficient). Since $f$ is well-defined on $2G$-cosets, this subspace is preserved by the oracle $U_f$, and we can think of its restriction to that subspace as $ U_{\bar{f}} $.
	
	Therefore, any instance of the minimal-label index-$2$ HSP can be reduced to the case of $ G = \left( \Ztwo \right)^n $; and the reduction can be performed without any oracle queries, as the above construction is independent of the hidden subgroup $H$. As we have shown above, the minimal-label index-$2$ HSP with $ G = \left( \Ztwo \right)^n $ is equivalent to the Bernstein--Vazirani problem, except that $ \bar{f} $ may fail to be a group homomorphism; but since $2$ is prime, the corresponding $\iota$ satisfies the Fourier-support condition automatically. Hence, our algorithm for the minimal-label index-$2$ HSP can be reduced to the Bernstein--Vazirani algorithm. While this reduction requires no oracle queries, it does entail the following computational steps:
	\begin{itemize}
		\item Constructing an explicit isomorphism $ \phi : G/2G \xrightarrow{\sim} \left( \Ztwo \right)^n $---this can be done classically and ``offline'', without any access to the oracle;
		
		\item Using $ \phi $ to organize the coset states $ \ket{g+2G} $ into qubits, and implementing Hadamard gates acting on those qubits;
		
		\item Mapping the measured character $ \rho \in \widehat{ \left( \Ztwo \right)^n} $ to a character of $ G/2G $, using the pullback $ \phi^* : \widehat{ \left( \Ztwo \right)^n} \xrightarrow{\sim} \widehat{G/2G} $ (defined via $ \rho \mapsto  \rho \circ \phi$);
		
		\item Converting $ \phi^* (\rho) \in \widehat{G/2G} $ back into a description of $H$ (the union of all $2G$-cosets $ c \in \ker \phi^* (\rho) \le G/2G $).
	\end{itemize}
	
	A similar reasoning holds for any $q$. Let $ qa \coloneqq \underbrace{a + \ldots + a}_{q \; \mathrm{times}} $, and define the subgroup $qG \coloneqq \{ qg : g \in G \}$. For any index-$q$ subgroup $H$, $ G/H $ has $q$ elements, so the order of every element in $ G/H $ must divide $q$; and if $H$ has index $1$ then $G/H$ is the trivial group, and the order of its element is $1$, which divides $q$. Thus, for any $g \in G$ we have $ \pi \left( q g \right) = q \pi \left( g \right) = 0 \in G/H $, which shows that $ qG \le \ker \pi = H $. Now, if $q$ is \emph{prime}, the quotient $ G / qG \cong \left( \Zq \right)^n $ for some $n$; and our algorithm from \Cref{sec:idx2HSP} solves the problem with a single query. The QFT we need is just the $n$-tensor product of the standard QFT over $ \Zq $.
	
	However, if $q$ is not a prime, then the quotient $ G / qG $ can have a more complicated form. Let $G$ have the primary decomposition $	G \cong \mathbb{Z} / r_1 \mathbb{Z} \oplus \cdots \oplus \mathbb{Z} / r_t \mathbb{Z} $, where $r_1, \ldots, r_t$ are prime powers. Thus, an arbitrary element $g \in G$ can be written uniquely as $ g = g_1 + \ldots + g_t $ for $ g_i \in \mathbb{Z} / r_i \mathbb{Z} $; and an arbitrary element in $qG$ has the form $ qg = q \left( g_1 + \ldots + g_t \right) = q g_1 + \ldots + q g_t $. This shows that $ qG \cong q \left( \mathbb{Z} / r_1 \mathbb{Z} \right) \oplus \cdots \oplus q \left( \mathbb{Z} / r_t \mathbb{Z} \right) $, hence the quotient is
	\begin{equation}
		 G / qG \cong \frac{\mathbb{Z} / r_1 \mathbb{Z}}{q \left( \mathbb{Z} / r_1 \mathbb{Z} \right)} \oplus \cdots \oplus \frac{\mathbb{Z} / r_t \mathbb{Z}}{q \left( \mathbb{Z} / r_t \mathbb{Z} \right)} .
	\end{equation}
	Now, for each $i$ we have $ \frac{\mathbb{Z} / r_i \mathbb{Z}}{q \left( \mathbb{Z} / r_i \mathbb{Z} \right)} \cong \mathbb{Z} / \mathrm{gcd} \left( q, r_i \right) \mathbb{Z} $. For prime $q$, $\mathrm{gcd} \left( q, r_i \right)$ can only be $1$ or $q$, so the only surviving factors are $ \mathbb{Z} / q \mathbb{Z} $; but for an arbitrary $q$, the gcd can be any number that divides $q$, so in general we get a decomposition of the form
	\begin{equation}
		G / qG \cong \bigoplus_{a \ge 2 \, : \, a \vert q} \left( \mathbb{Z} / a \mathbb{Z} \right)^{n_a} ,
	\end{equation}
	for some non-negative integers $n_a$.
	
	\subsection{Generalizations of Bernstein--Vazirani}
	Using the above description of the BV problem and algorithm, it seems natural to seek a generalization. By the preceding discussion, the minimal-label index-$q$ HSP comprises one such generalization.
	
	Zhandry~\cite{zhandry2015quantum} generalizes BV even further, considering functions $f : S \rightarrow G$ whose domain $S$ is an arbitrary set and codomain $G$ an arbitrary finite abelian group. Note the assumptions on the domain and codomain are the same as in the version of Deutsch--Jozsa--H{\o}yer discussed in \Cref{sec:DJH}; however, here we are generally not promised anything else about the function $f$ (in contrast to Deutsch--Jozsa--H{\o}yer, where $f$ is either constant or balanced).
	The set of such functions has the structure of an abelian group (by pointwise addition), and the shift oracle corresponds to a representation of this group. Thus, the oracle $ U_f $ represents an \textit{unknown} element $ f $ of the group $ \mathrm{Mor} \left( S, G \right) $ via a \textit{known} representation; and we are tasked with finding $f$. More generally, we may be tasked with finding the coset to which $f$ belongs, with respect to some (known) subgroup $ H \le \mathrm{Mor} \left( S, G \right) $.
	
	Thus, this coset identification problem is a vast generalization of BV. However, the optimal algorithm presented by Zhandry may not use a QFT in the general case where $S$ is not a group; in fact, a single-query algorithm that always succeeds may not exist. The relation with BV is mentioned in~\cite{Copeland2021quantumquery}, which generalizes Zhandry's paper to the case of non-abelian $G$.
	
	Let us emphasize the key distinctions between Zhandry's coset identification problem and the HSP. The HSP involves a function $f$ whose domain is a group, while in coset identification the codomain of $f$ is a group. In the HSP, $f$ is promised to hide some subgroup $ H $ of its domain (i.e. $f$ factors through a quotient map---see \eqref{triangle}). Therefore, the group structure of the domain $G$ of $f$ plays a key role in the problem formulation; and quantum algorithms for the HSP utilize this structure to solve the problem efficiently (e.g. Shor--Kitaev for abelian $G$, or our algorithm from \Cref{sec:idx2HSP} for the special case of index-$q$ HSP). In contrast, in coset identification $f$ is an arbitrary function $ S \rightarrow G $, and we think of $f$ as an element of the group $ \mathrm{Mor} \left( S, G \right) $. Here we are given a subgroup $H \le \mathrm{Mor} \left( S, G \right) $ and are tasked with finding the coset to which $f$ belongs. Hence, the main algebraic structure underlying the problem (and algorithm) is the group structure of $ \mathrm{Mor} \left( S, G \right) $, as well as the unitary representation $U_f$. The more general version of coset identification~\cite{Copeland2021quantumquery} distills this structure: the only assumption is that the oracles form a unitary representation of \textit{some} group. This group is not necessarily given as $ \mathrm{Mor} \left( S, G \right) $ for some group $G$, hence the oracles are not assumed to be shift oracles (in fact, they do not necessarily represent functions).
	
	\section{Conclusion}\label{sec:conclusion}
	In this work we revisited the role of algebraic structure in two of the earliest quantum algorithms. By reformulating the Deutsch--Jozsa problem independently of any group structure, we clarified when the appearance of the Hadamard (and more generally, the quantum Fourier) transform is intrinsic to the problem and when it is incidental. This perspective allowed us to place several known generalizations under a common framework; to introduce the index-$q$ hidden subgroup problem and solve it in a single query; and to show that the Bernstein--Vazirani problem forms a special case of the minimal-label index-$q$ HSP.
	
	More broadly, our results highlight that the algebraic structure, provided as a promise, gives rise to query-efficient quantum algorithms. By distinguishing between the structural features of a problem and the circuit-level details of its implementation, we gain a clearer view of which aspects of quantum algorithms are essential and which are interchangeable. This not only sharpens our understanding of existing algorithms, but also suggests new avenues for generalization, where identifying the right algebraic structure could lead to efficient quantum solutions for classes of problems not yet known to admit them.
	
	Future work could aim to characterize exact one-query solvability for output alphabets larger than $ \left[ G : H \right] $. Moreover, it is of interest to quantify bounded-error performance in terms of the largest cyclic factor of $G/H$. Finally, one may extend the analysis beyond the finite setting, adapting the framework to locally compact (infinite) groups with open finite-index subgroups.
	
	\section*{Acknowledgments}
	We are grateful to Greg Kuperberg for helpful discussions.
	AT and EC were partially supported by the European Union's Horizon Europe research and innovation programme under grant agreement No. 101178170 and by the Israel Science Foundation under grant agreement No. 2208/24. YO was supported by the Israeli Science Foundation Excellence Center, the US--Israel Binational Science Foundation, the Israel Ministry of Science.
	This work was supported by the Israeli Council for Higher Education through	``QERNEL'' and the ``Interdisciplinary Center for the Theory of Quantum Computing''.
	
	\section*{Data Availability Statement} 
	No data were generated or analyzed in this work.

\end{document}